\documentstyle[aps,prl,multicol,graphicx]{revtex}
\begin{document}
\draft \preprint{draft}
\title{Novel Phases in the Field Induced Spin Density Wave State in
(TMTSF)$_2$PF$_6$}
\author{A.\ V.\ Kornilov$^a$, V.\ M.\ Pudalov$^a$, Y.\ Kitaoka$^b$, K.\
Ishida$^b$, T.\ Mito$^b$, J.\ S.\ Brooks$^c$, J.\ S.\ Qualls$^c$,
J.\ A.\ A.\ J.\ Perenboom$^d$, N.\ Tateiwa$^e$, T.\ C.\ Kobayashi$^e$}
\address{$^{a)}$P.\ N.\ Lebedev Physics Institute, 117924 Moscow, Russia.\\
$^{b)}$Division of Material Physics, School of Engineering
Science, Osaka University,
Toyonaka, Osaka 560-8531, Japan\\
$^{c)}$ National High Magnetic Field Laboratory, Florida State
University,
Talahasse FL 32310 USA\\
$^{d)}$ High Field Magnet
Laboratory, University of Nijmegen, NL-6525 ED Nijmegen, The
Netherlands\\
$^{e}$ Research Center for Materials Science at Extreme Condition, Osaka University,
 Toyonaka, Osaka 560-8531, Japan}

\date{\today}
\maketitle

\begin{abstract}
Magnetoresistance measurements on the quasi one-dimensional
organic conductor (TMTSF)$_2$PF$_6$ performed in magnetic fields $B$
up to 16\,T, temperatures $T$ down to 0.12\,K and under
pressures $P$ up to 14\,kbar have revealed new phases on its $P-B-T$
phase diagram. We found a new boundary  which subdivides
the field induced spin density wave (FISDW) phase  diagram
into two regions. We showed that a low-temperature region
of the FISDW diagram is  characterized by a hysteresis behavior
typical for the first order transitions,
as observed in a number of studies.
In contrast to the common believe, in
high temperature region of the FISDW phase diagram,
the hysteresis and, hence, the first order transitions
were found to disappear.
Nevertheless, sharp changes in the resistivity slope
are observed both in the low and high temperature domains
indicating that the cascade of transitions between different
subphases exists over all  range of the FISDW state.
We also
found that the temperature dependence of the resistance (at a
constant $B$) changes sign
 at about
the same boundary. We compare these
results with recent theoretical  models.

\end{abstract}

\begin{multicols}{2}
Layered organic compounds tetramethiltetraselenafulvalene
(TMTSF)$_2X$, where the anion $X$ is ClO$_4$, AsF$_6$, PF$_6$ etc
are unique material systems with a very rich phase diagram (for a
review see
Refs.~\cite{gorkov&lebed,gorkov84,ishiguro98,chaikin96}).
Conduction in this materials is highly anisotropic, with ratio of
the components $\sigma_{xx} : \sigma_{yy} : \sigma_{zz} \sim 10^5
: 10^3 : 1$.
At ambient pressure,  below temperature of
12\,K, the PF$_6$-compound becomes dielectric with antiferromagnetic spin
ordering in the spin density wave state. As pressure increases,
the temperature of the antiferromagnetic ordering decreases  and
at about $P=6$\,kbar a superconducting state sets in. Magnetic
field $B$ applied in the least conducting direction $z$, first
quenches the superconducting state and, further  induces a
cascade of
phase transitions between FISDW states accompanied by the quantum
Hall effect \cite{hannahs89,cooper89}.

A so called `standard'  model  was suggested
\cite{gorkov&lebed,gorkov84,chaikin85} to explain the metal-SDW
transition in magnetic field. Later it was developed into a
`Quantized Nesting Model' in Refs.
\cite{heritier84,lebed85,maki86,yamaji86} to describe a cascade
of the first order transitions between different FISDW sub-phases.
According to this model,  electrons condense in the
SDW state whose period determines a nesting vector in the
momentum space. Under the assumption of the electron-hole symmetry,
$x-$component of the nesting vector
$Q_x$
is quantized as \cite{heritier84,ishiguro98}
\begin{equation}
Q_x = 2k_F - N\frac{eBb}{h},
\end{equation}
where $k_F$ is Fermi wave vector,
$b$ is the size of the elementary cell in $y-$ direction, and $N$
is an integer. As magnetic field varies, $N$ changes by an
integer, causing step-like changes in the nesting vector, which
result in the sequence of the first-order phase transitions.

According to the recent analysis  \cite{lebed00}, however  the
electron-hole symmetry in the SDW state is not fulfilled  unless
$N=0$ in Eq. (1). As a result, (i) the nesting vector
is not strictly quantized and (ii) the step-like changes in the nesting
vector may disappear above a certain temperature transforming into
oscillations. Correspondingly, as temperature increases, the
first order transitions with $\Delta N\approx  1$ may disappear,
whereas FISDW state still persists. Thus, in contrast to the
`Quantized Nesting Model' which predicts the first order phase
transition to exist over the whole range of temperatures where
FISDW develops, the `novel model'  predicts the
first order phase transitions may disappear above a certain
temperature, $T_0$. The latter possibility
depends on the  parameter
$\hbar \omega_c/(2\pi k_BT_0)$  \cite{lebed00},
where $\omega_c$ is the cyclotron frequency.

In order to verify  the theoretical predictions of the  two
models above, we studied temperature dependence of the
magnetoresistance in (TMTSF)$_2$PF$_6$ at various pressures.
Specifically, we measured, at different pressures, a temperature
evolution of the hysteresis intrinsic to magnetoresistance traces
of $R(B)$. We observed
that the hysteresis indeed disappears above a temperature $T_0$
whereas FISDW state still persists. We found such behavior to
manifest itself over the whole explored range of the existence of
the FISDW. According to our results, the total $P-B-T$ phase
diagram of the  FISDW state can be subdivided into two domains,
the `low-$T$' domain where the first order phase transitions
between FISDW sub-phases take place, and the `high-$T$' domain
where the transitions between the FISDW states do not exhibit
first order behavior. This observation is in agreement with the
`novel model'; in the latter case, the `low $T$-phase'  is
treated as a `quantum FISDW' state with step-like changes in the
nesting vector, whereas the `high $T$-phase' is treated as the
`semi-classical FISDW' state where the nesting vector oscillates.
We also found that as temperature decreases and crosses the
boundary between the two domains, the behavior of the resistivity
changes qualitatively, from ordinary insulating-like
($dR/dT<0$) through the `high-$T$'-domain, to the
metallic-like  ($dR/dT >0$) through the `low-$T$'-domain. Around
the $T=T_0$-boundary, $R(T)$ exhibits a maximum.

{\it Experimental}. Three samples (of a typical size $2\times
0.8\times 0.3$\,mm$^3$) were grown by a conventional
electrochemical technique.
Measurements were made using  either four Ohmic contacts
formed at the $a-b$ plane or eight contacts at two  $a-c$ planes;
in all cases $25\mu$m Pt-wires were attached by a graphite paint to the sample
 along the most conducting direction $a$. The sample and a
manganin pressure gauge were inserted  into a Teflon cylinder
placed inside a nonmagnetic 18\,mm o.d. pressure cell
\cite{cylinder_cell} filled with Si-organic pressure transmitting
liquid. The cell was mounted inside the liquid He$^4$,  He$^3$, or
He$^3$/He$^4$
chamber,  in a bore of a 16\,T superconducting magnet. For all
measurements, the magnetic field was applied along the least
conducting direction, $z$, of the crystal. Sample resistance was measured by four
probe ac technique at 132\,Hz frequency, with  current 1-4 $\mu$A to
avoid nonlinear effects. The out-of phase component of the measured
voltage was found to be  negligible in
all measurements, indicating Ohmic contacts to the sample.

The sample  temperature was varied slowly, at a rate less than
0.25 K/min in order to avoid breaking the sample. The measured
changes in the sample resistance were fully reproducible during
the full run of measurements including temperature sweeps; this
indicated  that the sample quality did not change.
The magnetoresistance was measured either at a constant $T$ and
varying  magnetic field $B$, or at a constant $B$ and varying
$T$. Sample temperature was determined by RuO$_2$ resistance
thermometer with a pre-calibrated magnetoresistance.
Measurements were done in magnetic fields up to 16\,T and for
temperatures in the range from 1.4 to 30\,K (mainly) and down to
$0.12$\,K (partly). The most detailed results were obtained for
pressures 7, 8, 10 and 14\,kbar.

Figure 1 shows magnetoresistance traces measured (a) at $P=
10$\,kbar in the temperature range 0.6-4.2\,K  and (b) at 8\,kbar,
 0.12\,K. In agreement with earlier observations
\cite{greene82}, when magnetic field exceeds the
critical value (which is 0.16\,T in our case), the superconductivity is quenched and
the sample resistance starts gradually increasing. Further, this
smooth dependence transforms into step-like changes in $R$. As
temperature decreases, the step-like changes become steeper and
appear at progressively lower fields. This behaviour is
also consistent with earlier observations
\cite{chaikin83,ribault83,hannahs89,cooper89} and is
interpreted as transitions between different sub-phases in FISDW
\cite{kwak,cooper89,hannahs89}. This interpretation is further
supported by the hysteresis between $R(B)$ traces for the field
ramping up and down, which is clearly seen in Fig.~1. The
hysteresis is also consistent with earlier observations
\cite{cooper89}
and signals the onset of the first order phase transitions.
\begin{figure}
\begin{center}
\includegraphics[angle=0,width=3.0in]{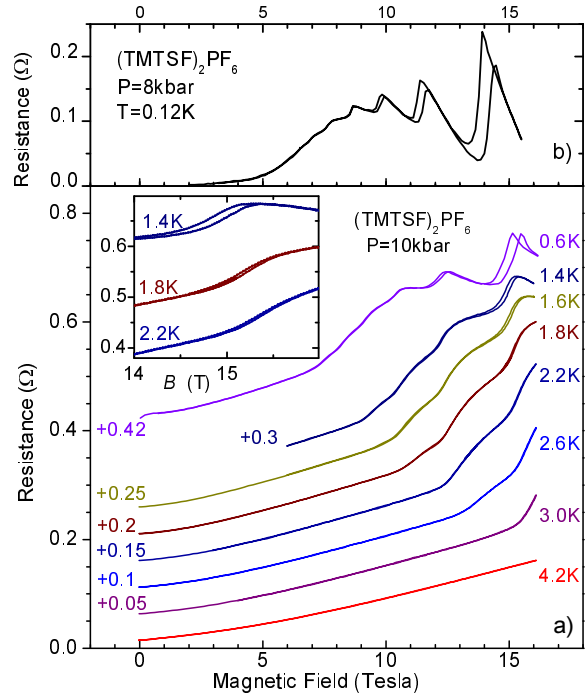}
\begin{minipage}{3.2in}
\vspace{0.1in} \caption{Magnetoresistance $R_{xx}$ vs magnetic
field $B_z$: a) For $P=10$\,kbar and for eight temperatures
(indicated for each curve). The curves are shifted vertically, for clarity
(the offset
values are indicated on the left side). The
inset magnifies the temperature evolution of the hysteresis regions
of $R(B)$-curves near $B=15$\,T. b) For $P=8$\,kbar and $T=0.12$\,K.}
\label{fig1}
\end{minipage}
\end{center}
\end{figure}
\vspace{-0.1in}

As temperature increases, the hysteresis weakens and tends to
disappear
as illustrated in the
inset to Fig.~1\,a. Nevertheless, the steps in $R(B)$ persist to
higher temperatures, being therefore non- or at least partly
correlated with the hysteresis. In order to quantify the
hysteresis strength, we calculated the maximal width of the
hysteresis loop $\delta B$ for each  curve and plotted it
in Fig.~2  as a function of temperature; in this
determination, the $L/R$ time constant of the magnet was
carefully measured and taken into account.

For three groups of the data in Fig.~2 the  hysteresis width
decreases linearly with temperature and
vanishes at a certain temperature $T_0$;
above  $T=T_0$ it remains  equal  to zero.
The falling part of these dependences were fitted with linear curves (solid lines),
which appear to have the same slope.
We plotted linear curves with the same slope
through other single data points (dashed lines) in order to estimate
$T_0$ for all transitions at different pressure values.

Measurements at two other pressures, 7 and 14\,kbar  have
shown qualitatively similar results. At $P=7$\,kbar the steps
(transitions) shift to lower
fields and persist up to higher temperatures. The
hysteresis, $\delta B$, is bigger than that at 8 and
10\,kbar and disappears at slightly  higher temperature. At
$P=14$\,kbar, the trend is opposite: $T_0$ becomes lower  than
that for 10\,kbar.

\begin{figure}
\begin{center}
\includegraphics[angle=0,width=3.0in]{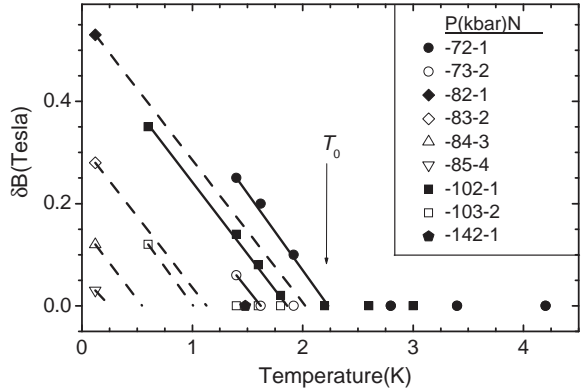}
\begin{minipage}{3.2in}
\vspace{0.1in}
\caption{Hysteresis width vs
temperature for several pressures.
Lines are the guide to the eye.
Dashed lines show an anticipated behavior in cases where only single data point was taken.
The table shows pressure values $P$ (in kbar) and the sub-phase numbers $N$ between which the
transition takes place. Vertical arrow depicts $T_0$ for one of
the transitions, $N=2\longleftrightarrow 1$ at $P=7$\,kbar.}
\label{fig2}
\end{minipage}
\end{center}
\end{figure}
\vspace{-0.1in}

 The above three features, (i) the existence of
the hysteresis in $R(B)$ at low temperature, (ii) its
disappearance above a certain temperature $T_0$ and (iii) the
persistence of the steps in $R(B)$ to temperatures higher than
$T_0$, are observed  in our experiments for
several transitions (see Figs.~1, 2).
It seems  likely that these features are generic also to other
transitions in the FISDW part of the phase diagram and that the
hysteresis for higher $N$-values was not observed in our
measurements just because  $T_0$ for these transitions is lower
than our lowest accessible temperature, 1.4\,K (for the majority
of measurements).

\begin{figure}
\begin{center}
\includegraphics[angle=0,width=2.8in]{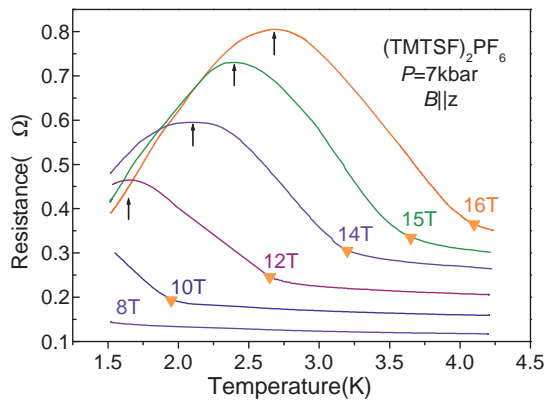}
\begin{minipage}{3.2in}
\vspace{0.1in}
\caption{Temperature dependences of the resistance
for six different values of the magnetic field at pressure
$P=7$\,kbar. Triangles depict the onset of the transitions.
Arrows point the maximum in each curve at $T_{\rm max}$.}
\label{fig3}
\end{minipage}
\end{center}
\end{figure}
\vspace{-0.1in}

Figure 3 represents the results of
temperature sweeps taken at six fixed magnetic fields for $P=7$\,kbar.
 Starting from high temperatures, the resistance
increases as $T$ decreases, then passes through a maximum at a
certain temperature $T_{\rm max}$ and  further decreases towards
low temperatures. Triangles mark the onset of the phase transition at each curve.
The similar $T-$dependences measured at $P=10$ and 14\,kbar were qualitatively similar
to those shown in Fig.~3 but shifted to lower temperatures.

$B-T$ phase diagram in Fig.~4 summarizes the  results of all measurements,
at $P=7$\,kbar (the main panel) and at $P=10$\,kbar (the inset).
The closed squares depict the onsets of the steps in $R(B)$ obtained from magnetic field sweeps
at fixed temperatures and triangles  are for the  temperature
sweeps $R(T)$ at fixed field.
In addition to the data taken directly,
the open squares
show the lower temperature data, $T=0.12$\,K,
taken at $P=8$\,kbar which has been
recalculated to correspond to the data at $P=7$\,kbar (main panel)
and to 10\,kbar (inset).
In this procedure, the data for 8\,kbar
were shifted in magnetic field according to the
pressure coefficient $d (B^{-1})/dP = - 0.015$\,T$^{-1}$kbar$^{-1}$
which we determined from the higher temperature data at
$T=1.4$\,K for $P=7$, 10, and 14\,kbar. The hysteresis width is
 obtained from Fig.~2 and is depicted
by the split lines.
\begin{figure}
\begin{center}
\includegraphics[angle=0,width=3.0in]{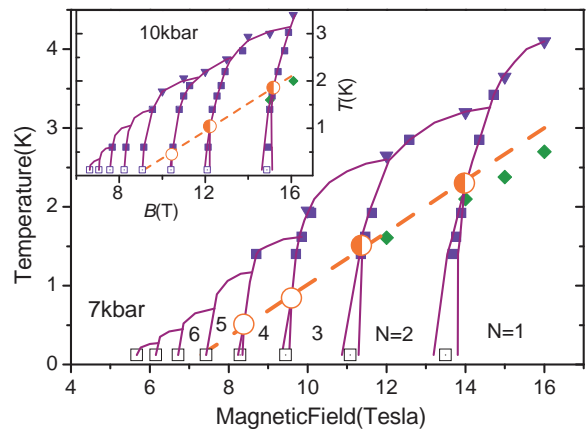}
\begin{minipage}{3.2in}
\vspace{0.1in} \caption{$B-T$  -- phase diagram
for   $P=7$\,kbar (main panel), and for
$P=10$\,kbar (inset). Solid squares (triangles) correspond to the onset of
the steps in $R(B)$ ($R(T)$) curves measured directly at the specified pressure
($P=7$ and 10\,kbar, correspondingly).
Open squares are recalculated from the data taken at $P=8$\,kbar.
Circles denote $T_0$: semi-closed symbols are for the data
measured directly at the specified pressure,
open ones are
recalculated from the data measured at $P=8$\,kbar.
Diamonds designate the temperature $T_{\rm max}$ of
maxima in $R(T)$ measured directly at the specified pressure
(see Fig.~3 for $P=7$\,kbar).
Split lines of the phase diagram correspond to the hysteresis
width;
all other lines are guide to the eye. }
\label{fig4}
\end{minipage}
\end{center}
\end{figure}
\vspace{-0.1in}
In general, the $P-B-T$ phase diagram in Fig.~4 is qualitatively
similar to that
previously reported \cite{cooper89,hannahs89},
but, in addition, displays  the boundaries of hysteresis domains
vs temperature.
The hysteresis domains for different transitions
collapse above
$T=T_0$; this was
 determined for
seven transitions and the corresponding points are denoted
with open circles. The separatrix points $T_0$ thus
subdivide the phase boundaries into the two regions: the low
temperature domain ($T<T_0$) of the hysteretic behaviour and the
high temperature domain ($T>T_0$) where the FISDW transitions
develop without a hysteresis. The disappearance of the hysteresis
with rising temperature at one fixed pressure was mentioned
earlier \cite{cooper89}
 but to the best of our knowledge no studies of this effect
followed.
The subdivison of the phase diagram is qualitatively consistent with the `novel model'
for the FISDW \cite{lebed00}, where the low-temperature domain
corresponds to the `quantum FISDW' sub-phase and the high
temperature domain corresponds to the `semiclassical FISDW'. According
to the model, in the former sub-phase  the transitions between
different phases take place with jumps in the nesting vector,
are of  the first order, and are accompanied by hysteresis of various
physical quantities. In the latter sub-phase, the transitions
between different phases take place without any jumps in the
nesting vector,
and are therefore not first order phase transitions.

The
$B-T$ phase diagram on the main panel in Fig.~4 is plotted for pressure of 7\,kbar.
For $P=10$\,kbar (inset), the phase diagram looks
qualitatively similar but the new phase boundary is shifted to
lower temperature by about (0.5 - 1)\,K.

The dashed line in Fig.~4 connects the $T_0-$values
(circles). In order to check whether the dashed line
splits not only the phase boundaries but the overall
$B-T$ parameter space, we plotted onto the same phase diagram
 the $B,T$-coordinates of the maxima on $R(T)$ curves from
 $T-$sweeps similar to that shown in Fig.~3.

The maxima in $R(T)$ dependences in the FISDW regime were observed
earlier \cite{kang92,vuletic} and were associated  with
a non-linear temperature dependence of the $\sigma_{xy}$  caused
by the transition from the QHE to metallic regime
\cite{kang92,yakovenko,vuletic}.
In this
interpretation, the maxima should correspond to the onset of the
QHE regime. On the other hand, the hysteresis behaviour if
associated with jumps in  the nesting vector \cite{lebed00},
should manifest
in the QHE regime only. Therefore, $T_{max}$ is
expected to be $\leq T_0$.  It is surprising that in our
case $T_{\rm max}$
almost coincide with $T_0$ for different  pressures,
even though $T_0$ varies with pressure.

{\it To summarize},
from studies of the temperature  and magnetic field
dependences of the resistivity of the quasi-1D organic conductor, we
found that its $P-B-T$ phase diagram splits in two domains, where
the transitions between different FISDW states take place
(i) with a hysteresis
as the first order phase transitions (for low
temperatures),
and (ii) without hysteresis (for high temperatures).
This result is not expected within the `Quantized nesting model'
and is consistent
with  the recent suggestion by Lebed \cite{lebed00} that the period  of the
spin
structure in FISDW state can be either partially quantized or
not  quantized at all.
We experimentally plotted the new phase
boundary where such behavior occurs. We found that the $P-B-T$
boundary separating the two domains almost coincides  with a line
at which the $dR/dT$ changes sign.

\vspace{-0.1in}
\acknowledgements
\vspace{-0.1in}
A.\ V.\ K.\ and V.\ M.\ P. are grateful to A.\ G.\ Lebed and V.\ M.\ Yakovenko
for fruitful discussions. The work was partially supported by INTAS, RFBR,
NATO, NSF, NWO, Programs ``Statistical physics'',
``Integration'', ``The State support of the leading scientific
schools'', and COE Research in Grant-in-Aid for Scientific Research, Japan.

\end{multicols}

\end{document}